\documentclass[%
 reprint,
superscriptaddress,
 amsmath,amssymb,
 aps,
]{revtex4-1}

\usepackage{graphicx}% Include figure files
\usepackage{dcolumn}% Align table columns on decimal point
\usepackage{bm}% bold math
\usepackage{xcolor}
\usepackage{soul} % For strikethrough text
\usepackage{hyperref}
\hypersetup{colorlinks=true,citecolor={blue},linkcolor={blue},urlcolor={blue}} % To make links blue
\usepackage{upgreek}
\renewcommand{\mu}{\upmu} % Make \mu symbol for microns
\newcommand{\um}{\mu\textrm{m}}			% microns
\newcommand{\iium}{\mu\textrm{m}^{-2}} % inverse microns notation
\newcommand{\ips}{\textrm{ps}^{-1}} % inverse picosecond notation

\begin{document}

\preprint{APS/123-QED}

\title{Optically trapped polariton condensates as semiclassical time crystals}% Force line breaks with \\
%\thanks{A footnote to the article title}%

\author{A. V. Nalitov}
\affiliation{Science Institute, University of Iceland, Dunhagi 3, IS-107, Reykjavik, Iceland}
\affiliation{ITMO University, St. Petersburg 197101, Russia}

\author{H. Sigurdsson}
\affiliation{Science Institute, University of Iceland, Dunhagi 3, IS-107, Reykjavik, Iceland}

\author{S. Morina}
\affiliation{Science Institute, University of Iceland, Dunhagi 3, IS-107, Reykjavik, Iceland}

\author{Y. S. Krivosenko}
\affiliation{ITMO University, St. Petersburg 197101, Russia}

\author{I. V. Iorsh}
\affiliation{ITMO University, St. Petersburg 197101, Russia}

\author{Y. G. Rubo}
\affiliation{Instituto de Energ\'{\i}as Renovables, Universidad Nacional Aut\'onoma de M\'exico, Temixco, Morelos 62580, Mexico}

\author{A. V. Kavokin}
\affiliation{International Center for Polaritonics, Westlake University, No.18, Shilongshan Road, Cloud Town, Xihu District, Hangzhou, China}
\affiliation{CNR-SPIN, Viale del Politecnico 1, I-00133, Rome, Italy}
\affiliation{Spin Optics Laboratory, St. Petersburg State University, St. Petersburg, 198504, Russia}
\affiliation{Russian Quantum Center, 100 Novaya Street, Skolkovo, Moscow Region 143025, Russia}

\author{I. A. Shelykh}
\affiliation{Science Institute, University of Iceland, Dunhagi 3, IS-107, Reykjavik, Iceland}
\affiliation{ITMO University, St. Petersburg 197101, Russia}

\date{\today}% It is always \today, today,

\begin{abstract}
We analyse nonequilibrium phase transitions in microcavity polariton condensates trapped in optically induced annular potentials.
We develop an analytic model for annular optical traps, which gives an intuitive interpretation for recent experimental observations on the polariton spatial mode switching with variation of the trap size.
In the vicinity of polariton lasing threshold we then develop a nonlinear mean-field model accounting for interactions and gain saturation, and identify several bifurcation scenarios leading to formation of high angular momentum quantum vortices.
For experimentally relevant parameters we predict the emergence of spatially and temporally ordered polariton condensates (time crystals), which can be witnessed by frequency combs in the polariton lasing spectrum or by direct time-resolved optical emission measurements.
In contrast to previous realizations, our polaritonic time crystal is spontaneously formed from an incoherent excitonic bath and does not inherit its frequency from any periodic driving field.
\end{abstract}

\pacs{Valid PACS appear here}% PACS, the Physics and Astronomy
                             % Classification Scheme.
%\keywords{Suggested keywords}%Use showkeys class option if keyword
                              %display desired
\maketitle

The idea of a time crystal, a state of matter characterized with discrete translation symmetry in both space and time, was recently proposed by Frank Wilczek \cite{Wilczek2012,Shapere2012}, shortly followed by the establishment that the absence of thermodynamic equilibrium is a necessary prerequisite for  its realization \cite{Bruno2013,Bruno2013a,Watanabe2015}.
Recently, two groups reported observation of time crystals in periodically driven discrete trapped ion systems \cite{Zhang2017} and dipolar interacting diamond impurities \cite{Choi2017}.
The observation of time translation symmetry breaking in time crystals extends the limits of relativistic analogy between space and time and thus has a huge fundamental significance.

The absence of thermodynamic equilibrium is naturally fulfilled for exciton-polariton condensates in microcavities, created by continuous incoherent optical or electric pumping \cite{Carusotto2013}, as the typical polariton thermalization time is longer then its lifetime.
Although Mermin-Wagner theorem forbids two-dimensional bosonic condensation with long-range order \cite{Hohenberg1967}, trapped cavity polaritons may macroscopically populate a size quantized single-particle state \cite{Balili2007}.
Among different polariton trapping schemes, such as mechanical strain \cite{Balili2007} or cavity etching \cite{Nguyen2013}, optically created traps have recently attracted significant attention due to extremely high tunability \cite{Gao2015,Askitopoulos2015a,Gao2018}. The optical confining potential stems from polariton repulsion off an inhomogeneous excitonic reservoir, typically generated by a spatially modulated light beam.
At the same time, the reservoir provides an inflow of polaritons to compensate for their decay, mainly governed by photon escape from the cavity, and, above a certain threshold density, supports a stationary condensate population \cite{Askitopoulos2013,Cristofolini2013}.
Nonequilibrium polariton condensates may thus occupy an excited trapped single-particle mode should the latter have higher net gain than the ground state.

Polariton-polariton interaction plays a crucial role here, as it can lift occasional degeneracy of the state, occupied by the driven condensate, and spontaneously break either translational \cite{Zhang2015}, spatial inversion \cite{Aleiner2012} or parity symmetry \cite{Ohadi2015}. Moreover, polariton-polariton interaction, supplemented with dissipative coupling, is sufficient for condensate stabilization in the weak lasing regime \cite{Aleiner2012}.
Finally, parametric polariton-polariton scattering out of the condensate may populate several energy levels, resulting in a multi-mode condensation \cite{Tosi2012}.

Polaritons repel through electron or hole Coulomb exchange interaction due to their excitonic component, governed by the Hopfield coefficient \cite{Glazov2009,Vladimirova2010}.
However, in the presence of a hot excitonic reservoir, condensed polaritons may also effectively attract due to local reservoir depletion \cite{Estrecho2018} and lattice heating \cite{Dominici2015}, resulting in condensate instability through self-localization \cite{Wouters2008}.
Destabilized resonantly driven polariton condensates may follow limit cycles and chaotic dynamics in both their density \cite{Solnyshkov2009} and polarisation \cite{Gavrilov2016,Gavrilov2018}.
The limit cycle behavior is also inherent to the weak lasing regime \cite{Rayanov2015}.

%Arbitrary lattices of coupled polariton condensates in optical traps present a peculiar platform for simulating interaction effects on bosonic many-body states.
%In particular, coupled condensate lattices exhibit ferromagnetic and antiferromagnetic phase \cite{Ohadi2016} and spin ordering \cite{Ohadi2016a}.
%Apart from forming steady states with stationary measurables, coupled driven polariton condensates follow limit cycle \cite{Rayanov2015} and chaotic Josephson oscillations of density \cite{Solnyshkov2009} and polarisation \cite{Gavrilov2016}. {\color{red}\it{(We are not considering a lattice of polaritons so I don't see why they should be mentioned. It would be better to introduce the reader here to results which have shown evidence of limit cycles in polariton systems. Not specifically lattices. - H)}}

In this Letter we focus on the properties of a single annular optical trap, where a condensate occupies a quasi-degenerate excited mode doublet.
The natural basis of modes in a rotational symmetric Hamiltonian are then the angular harmonics.
In such optical traps the formation of pinned and stable quantum vortices was recently predicted \cite{Yulin2016} and observed \cite{Manni2013,Gao2018}.
At the same time, polariton condensation into spatially ordered high angular momentum modes was observed in wider optical traps \cite{Dreismann2014b,Sun2018}.
We demonstrate the interrelation between these observations and point out that a new spatially and temporally ordered phase has been yet missed.

\emph{Two-mode model}. We describe the condensate with the two-dimensional ($\nabla^2 = \partial_x^2 + \partial_y^2$) complex Gross-Pitaevskii equation (GPE)
\begin{equation}  \label{eq_GPE}
\mathrm{i} { \partial \Psi \over \partial t} =
\left[ -{\nabla^2 \over 2m} + {n \over 2} \left(\alpha + \mathrm{i} \beta \right) + {\alpha_1 \over 2} \vert \Psi \vert^2 - \mathrm{i} {\Gamma \over 2} \right] \Psi,
\end{equation}
coupled to the semiclassical Boltzmann equation describing a reservoir of excitons sustaining the condensate
\begin{equation}
\label{eq_reservoir}
{\partial n \over \partial t} = \mathrm{P}\left( \mathbf{r} \right) - \left( \beta \vert \Psi \vert^2 + \gamma \right) n.
\end{equation}
Here $\Psi$ and $n$ are the condensate order parameter and the reservoir density, $\hbar m$ is the effective lower polariton mass, $\alpha$ and $\alpha_1$ are the interaction constants describing the polariton repulsion off the exciton density and polariton-polariton repulsion, $\beta$ governs the stimulated scattering from the reservoir into the condensate, $\Gamma$ and $\gamma$ are the polariton and exciton decay rates, and $\mathrm{P}(\mathbf{r})$ is the inhomogeneous reservoir pumping rate \cite{Wouters2007a}.
Assuming that the hot excitonic reservoir dynamics are fast on the timescale of the condensate evolution, its stationary density obtained from Eq.~\eqref{eq_reservoir} enters Eq.~\eqref{eq_GPE} as a function of $\vert \Psi \vert^2$, supplementing it with additional nonlinear terms \cite{Keeling2008}.

%\begin{figure}
%\includegraphics[width=1.0\linewidth]{fig1.jpg}
%\caption{(color online) Typical polariton condensate modes in optical traps. a) Ripple state in the harmonic non-Hermitian trap model. b) Petal state calculated with the rectangular trap model. Condensate, polariton current, and reservoir densities are plotted with green, blue, and white respectively. \label{Fig1}}
%\end{figure}

Small optical traps of sizes comparable to the characteristic reservoir variation scale are well approximated with the non-Hermitian harmonic oscillator model \cite{Askitopoulos2016a}, although its rotational symmetry is normally broken \cite{Nardin2010}.
As the potential created by an elliptic paraboloid reservoir density profile allows separating motion along the two principal axes, it explains the formation of "ripple" polariton modes in small optical traps \cite{Sun2018}.
%However, it fails to describe "petal" states {\color{red}\it{(Why does a Harmonic oscillator fail there? Angular harmonics are still good states. Are they more unstable or is the dominant gain always into the zero momentum ground state? - H)}}, similar to whispering gallery modes, with well defined high angular quantum number.
To describe wider traps we rather employ the non-Hermitian rotationally symmetric trap model, assuming a sharp edge between the reservoir $r>R$, and reservoir free region $r<R$, where $r$ is the radial coordinate of the planar microcavity system.
The former then corresponds to a uniform real potential and gain region provided by the homogeneous reservoir of density $n$.
Although realistic traps have an outer radius, we neglect polariton tunneling out of the trap governed by evanescent tails of confined state wavefunction into the barrier, keeping in mind that this approximation fails in the vicinity of confined state transitions to the continuum.

In the linear regime ($|\Psi|^2 \simeq 0$) and for a stationary reservoir ($dn/dt=0$), solutions to Eq.~\eqref{eq_GPE} can be written in the form $\Psi^\mathrm{n,m}(r,\varphi) = \exp(\mathrm{i m} \varphi) \Psi^\mathrm{n,m}(r)$, with $\mathrm{m}$ and $\mathrm{n}$ being the angular and the radial quantum numbers respectively.
%We point out that $m$ and $n$ are different from m and n.
%The former two correspond to the effective lower polariton mass in the parabolic regime of the dispersion the exciton reservoir density respectively. The latter are the angular and radial quantum numbers of the system solutions respectively.
The radial part $\Psi^\mathrm{n,m}(r)$ can be found in the two regions and has a piecewise defined form
\begin{equation} \label{def_WF}
\Psi^\mathrm{n,m}(r) = \left\lbrace \begin{matrix}
A^\mathrm{n,m} J_\mathrm{m} (r \sqrt{ 2 m E^\mathrm{n,m}}) , & r<R \\ B^\mathrm{n,m} K_\mathrm{m} (r \sqrt{ 2 m (U-E^\mathrm{n,m})}), & r>R
\end{matrix}\right.,
\end{equation}
where $J_\mathrm{m}$ and $K_\mathrm{m}$ are the analytic continuations of Bessel function of the first kind and Macdonald function of the second kind respectively, with their arguments being unambiguously defined as the principal square root values, and $U = \left( \alpha + \mathrm{i} \beta  \right) n/2 - \mathrm{i}\Gamma/2$.
Here the complex energies $E^\mathrm{n,m}$, as well as the normalization constants $A^\mathrm{n,m}$ and $B^\mathrm{n,m}$, are defined from equating the values and the first derivatives of the two wavefunction parts at the trap edge $r=R$, which yields the complex transcendental equation
\begin{equation} \label{eq_transcend}
-s_2 J_\mathrm{m}(s_1) K_{\mathrm{m}-1}^\prime(s_2) = s_1 K_\mathrm{m}(s_2) J_\mathrm{m}^\prime(s_1),
\end{equation}
where $s_1 = R \sqrt{2mE^\mathrm{n,m}}$, $s_2 = R \sqrt{2m(U-E^\mathrm{n,m})}$ and prime denotes differentiation.

\begin{figure}
\includegraphics[width=1.0\linewidth]{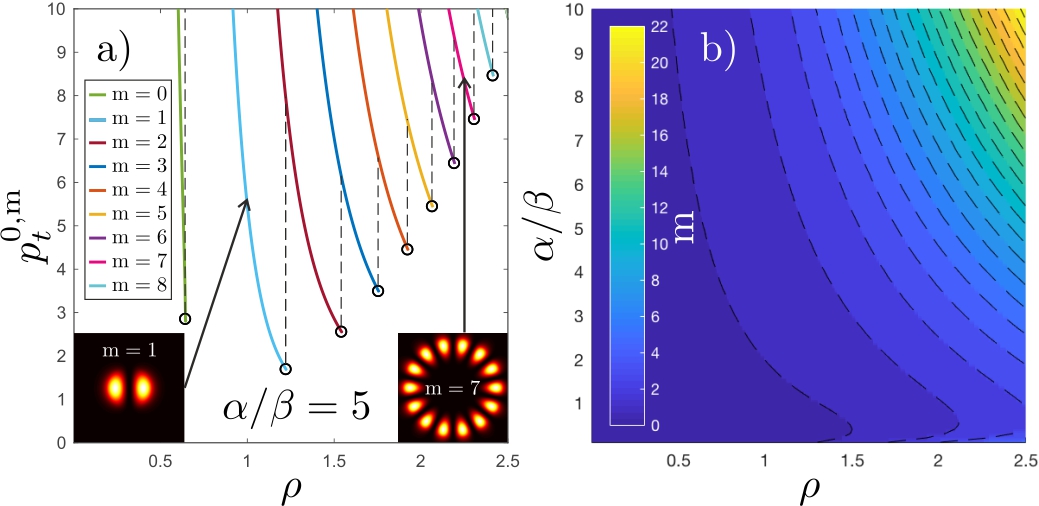}
\caption{(color online) Results of the linear non-Hermitian rectangular trap model. a) Reservoir pumping rate at the lasing threshold for states with $\mathrm{m}=0,1,\ldots,8$ as a function of the trap size $\rho = R \sqrt{2m\Gamma}$. Condensate switching points are marked with circles and dashed lines. b) Condensate angular quantum number $\mathrm{m}$ at the polariton lasing threshold as a function of the only two linear model parameters. \label{Fig1}}
\end{figure}

Polariton lasing threshold is defined as a point of gain-loss equilibrium of the linearized GPE \eqref{eq_GPE}, given by the condition $\mathrm{Im} \lbrace E^\mathrm{n,m} \rbrace = 0$, which occurs at a certain threshold reservoir density $n = n_t^\mathrm{n,m}$ in the region $r>R$.
Among the modes with a given angular momentum $\mathrm{m}$ the ground radial state $\mathrm{n}=0$ has the lowest threshold,
%{\color{red}(The optimal state corresponds to a single angular quantum number m due to the circular symmetry of the driving field (reservoir) but this state can be in a superposition of different radial states n. So the statement that the optimal state is n=0 is rather intuitive but not proven. Should this be mentioned? - HS)}
which explains why only this type of modes are observed at the threshold in large traps.
The physical reason behind this is the centrifugal force pushing a rotating condensate into the gain region.
Conveniently introduced dimensionless gain in the barrier $p_t^\mathrm{0,m} = \beta n_t^\mathrm{0,m}/\Gamma$, obtained by numerical solution of Eq.\eqref{eq_transcend}, is plotted in Fig.~\ref{Fig1}a for experimentally relevant relation $\alpha / \beta = 5$ as function of the dimensionless trap radius $\rho = R \sqrt{2m\Gamma}$.
%{\color{red} We note that $n_t^\mathrm{0,m}$ refers to the scalar value of the reservoir density for $r>R$.}
Every mode $\mathrm{m}$ has a critical trap radius of transition to the continuum, where $E_t^\mathrm{0,m} = \alpha n_t^\mathrm{0,m}/2$.
With increasing trap size $\rho$ the angular momentum of the polariton lasing threshold mode consequently increases through a cascade of successive switchings, as shown in Fig.~\ref{Fig1}b.
The angular momentum switching behaviour, as well as the superlinear oscillating pumping threshold dependence on the trap size, qualitatively reproduces the experimental data in Ref. \cite{Sun2018}.
%{\color{red}\it{(I replaced n=0 with n=1 in all quantities. - H)}}.

Any disorder in the pumping power distribution in the barrier and thus in the reservoir density leads to a splitting of both real and imaginary parts of energy for the two linear combinations of otherwise degenerate modes $\pm$$\mathrm{m}$.
A typical angular dependence of the lasing threshold mode density is therefore $\vert \Psi (\varphi) \vert^2 \propto 1 + \cos(2 \mathrm{m}\varphi)$, as shown in the insets of Fig. \ref{Fig1}a.

The role of the nonlinearity of Eq.~\eqref{eq_GPE} becomes increasingly important for pumping powers above the lasing threshold.
Assuming small trap asymmetry we neglect all modes except for the doublet $\pm \mathrm{m}$, corresponding to the condensation threshold.
Projecting Eq.~\eqref{eq_GPE} onto the basis $\Psi^{0,\pm\mathrm{m}}$ ($\Psi = \psi_+ \Psi^{0,\mathrm{m}} + \psi_- \Psi^{0,-\mathrm{m}}$) we have in the rotating wave frame:
\begin{align} \label{eq_proj_GPE}
\mathrm{i} { \mathrm{d} \psi_\pm \over \mathrm{d} t} = {1 \over 2} \left( \alpha + \mathrm{i} \beta \right) I_{cr} \left[ n_0 \psi_\pm +  {n_1 \mp \mathrm{i} n_2 \over 2} \psi_\mp \right] + \nonumber \\ {1 \over 2} \alpha_1 I_{cc} \left[ \vert \psi_\pm \vert^2 + 2 \vert \psi_\mp \vert^2 \right] \psi_\pm,
\end{align}
where $I_{cr} =2 \pi \int_R^{+ \infty} \vert \Psi^\mathrm{0,m} \vert^4 r dr$ is the condensate wavefunction overlap with the resevoir, $I_{cc} = 2 \pi \int_0^{+\infty} \vert \Psi^\mathrm{0,m} \vert^4 r \mathrm{d} r$ is the effective condensate overlap with itself, and the reservoir density angular harmonics, defined from the condition on the stationary reservoir density $\mathrm{d}n/\mathrm{d}t = 0$, being $n_0 =  (\mathrm{P} - I_{cr} p_t^{0,\mathrm{m}} \Gamma s ) / \gamma, \; n_1 = ( \delta \mathrm{P} - I_{cr} p_t^{0,\mathrm{m}} \Gamma s_x ) /\gamma, \; n_2 = -I_{cr} p_t^{0,\mathrm{m}} \Gamma s_y /\gamma$, and pseudospin components describing the condensate on the Bloch sphere defined as $s_x  = \mathrm{Re} \left\lbrace \psi_+^* \psi_- \right\rbrace$, $s_y = \mathrm{Im} \left\lbrace \psi_+^* \psi_- \right\rbrace$, $s_z = \left(\vert \psi_+ \vert^2 - \vert \psi_- \vert^2 \right)/2$, $\mathrm{P}$ is the angle independent part of the reservoir pumping rate variation from its threshold value $\mathrm{P}(\mathbf{r}) - \gamma n^\mathrm{0,m}$, while $\delta \mathrm{P}$ is the amplitude of the $\cos \left( 2 \mathrm{m} \varphi \right)$ reservoir density harmonic, assuming that the coordinates are chosen so that the corresponding $\sin \left( 2 \mathrm{m} \varphi \right)$ harmonic has a zero amplitude.

The evolution of the angular momentum pseudospin, obtained from Eq.~\eqref{eq_proj_GPE}, is then governed by the equation
\begin{equation} \label{eq_AM}
{\mathrm{d} \mathbf{S} \over \mathrm{d} \tau} = \left( P - S \right) \mathbf{S} + \left( \delta \mathbf{P} - \mathbf{S}_{\parallel}  \right) {S \over 2} + \left[\left( \varepsilon \delta \mathbf{P} - (\xi - \varepsilon) \mathbf{S}_\perp \right) \times \mathbf{S}\right],
 \end{equation}
with introduced dimensionless values $\tau = t \gamma$, $\mathbf{S} = \beta \Gamma I_{cr} p_t^\mathrm{0,m} \mathbf{s} / \gamma^2$, $\mathbf{S}_{\parallel}$ and $\mathbf{S}_\perp$ being the projections of $\mathbf{S}$ onto the $xy$ plane and the $z$ axis respectively, $P = \beta \mathrm{P}/\gamma$, and $\delta \mathbf{P} = \beta \delta \mathrm{P}/\gamma \mathbf{e}_x$ with $\mathbf{e}_x$ being the $x$ axis unitary vector.
Taking into account that $\alpha_1 \approx \vert X \vert^2 \alpha$ with $X$ being the excitonic Hopfield coefficient \cite{Brichkin2011}
%{\color{red}\it{(It is also written in Ref.[34] that $\alpha = 4 \alpha_1$ according to the Hartree-Fock theory, so I was wondering if it is more correct to write $\alpha_1 \approx \vert X \vert^2 \alpha/4$? - H)}}
, the two interaction parameters read
\begin{equation} \label{def_EX}
\varepsilon = {\alpha \over 2 \beta}, \; \xi = {\alpha \over \beta} {\vert X \vert^2 \over p_t^\mathrm{m}} {I_{cc} \over I_{cr}} {\gamma \over \Gamma}.
\end{equation}

The first two bracketed terms of Eq.~\eqref{eq_AM} represent gain-loss competition in the inhomogeneous pumping, while the last term describes absolute value conserving precession in the effective field.
The latter in turn has two contributions, one of them stemming from the pumping asymmetry, quantified by $\delta \mathrm{P}$, and the other one being the self-induced Larmor field \cite{Shelykh2004} with the effective interaction prefactor $\xi - \varepsilon$.
Depending on the relation between the effective polariton-exciton ($\varepsilon$) and polariton-polariton ($\xi$) interaction parameters the condensate is either in repulsive ($\xi>\varepsilon$) or in attractive ($\xi<\varepsilon$) regime.

\begin{figure}
\includegraphics[width=1.0\linewidth]{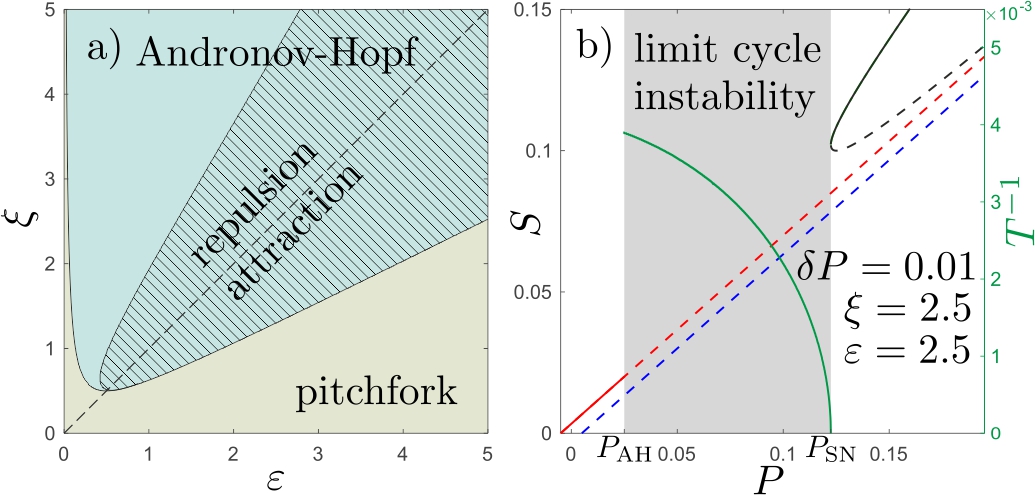}
\caption{(color online) a) Phase diagram of polariton condensate bifuractions. Areas of parameters corresponding to Andronov-Hopf(pitchfork) bifurcation type of the trivial stable solution branch $S^+(P)$ are shown in green(yellow) color respectively. The dashed line separates the repulsion (above) and attraction (below) regimes. The hatched blue area corresponds to limit cycle instabilities where only orbitally stable solutions exist. The unhatched blue area corresponds to bistability between the trivial and the symmetry breaking solution. b) Stationary condensate population dependence on pumping power. Stable(unstable) fixed points are plotted with solid(dashed) lines. Trivial $S^+(P)$ and $S^-(P)$ and the symmetry-breaking solutions are plotted in red, blue and black. The gray area highlights the instability range, and the frequency of limit cycles is plotted with green. \label{Fig2}}.
\end{figure}

Equation \eqref{eq_AM} has a pair of trivial stationary solutions, for which $S_y = S_z = 0$ and $S_x = \pm S$ corresponding to a petal state (e.g., shown in the inset of Fig.~\ref{Fig1}a) where the two counterrotating harmonics $\Psi^\mathrm{0,\pm m}$ are phase locked with $0$ and $\pi$ phase shifts.
The pumping power dependence of these trivial solutions reads $S^\pm (P) = \left(2 P \pm \delta P \right) / 3$.
There exists also another pair of stationary solutions, characterized by spontaneous parity symmetry breaking and nonzero $S_z$ and $S_y$ pseudospin components.
%\begin{equation}
%S_y = - {P-S \over \varepsilon \delta P} S_z, \; S_x = {P- {3 \over 2} S \over \varepsilon-\xi} {P-S \over \varepsilon \delta P} + {\varepsilon \delta P \over \varepsilon -\xi},
%\end{equation}
These two stationary pseudospins have the same absolute values and $S_x$ components, but the opposite signs of both $S_y$ and $S_z$, thus corresponding to the opposite directions of vorticity.
The transition between the two types of solutions explains the spontaneous formation of pinned quantum vortices in optical traps \cite{Askitopoulos2018}.
%{\color{red}\it{(Maybe move the above material to the supplemental and just mention that this vortex solution exists - H)}}.

The lower trivial branch $S^-(P)$ is unstable, while the higher $S^+(P)$ evolves either to a limit cycle at the Andronov-Hopf bifurcation point $P_{AH} = 5 \delta P /2$ or to a pair of symmetry breaking fixed points at the pitchfork bifurcation $P_{P1}$ provided $P_{P1} < P_{AH}$.
Fig.~\ref{Fig2}a shows the regions of two interaction parameters $\xi$ and $\varepsilon$, corresponding to these two scenarios of the trivial solution evolution with increasing pumping power $P$.
The vortex solutions, on the contrary, appear at a saddle-node bifurcation $P_{SN}$ in the condensate repulsion regime and at a pitchfork bifurcation $P_{P2}$ in the attraction regime.
See the Supplemental Material \footnote{Supplemental Material: [url]} for detailed derivation and analysis of Eq.\eqref{eq_AM}.

\begin{figure}
\includegraphics[width=1.0\linewidth]{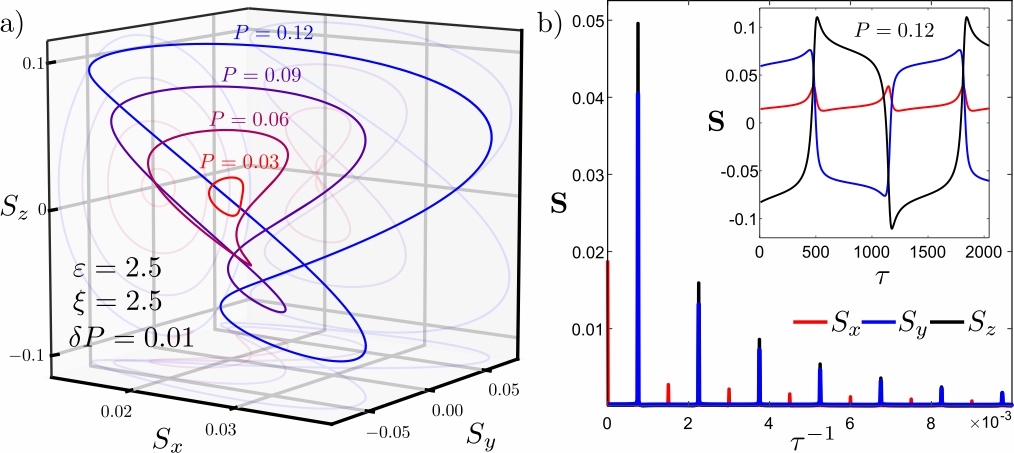}
\caption{(color online) Limit cycles of the condensate evolution. a) Pseudospin trajectories for pumping power spanning the instability range. b) Spectral and temporal (inset) dependence of the pseudospin projections in the anharmonic limit cycle regime, corresponding to the blue line in panel a). \label{Fig3}}
\end{figure}

In the following we discuss the nature of the periodic limit cycles, which is the only orbitally stable solution of Eq. \eqref{eq_proj_GPE} in the intermediate range of pumping powers between the stability regions of the trivial petal state and the parity breaking vortical state, as shown in Fig.~\ref{Fig2}b.
This range is only present in the Andronov-Hopf bifurcation scenario and if $P_{AH}<P_{SN}(P_{P2})$ in the repulsion(attraction) regime.
This condition is satisfied in a certain region of interaction parameters $\xi$ and $\varepsilon$, which is highlighted with hatching in Fig.~\ref{Fig2}a.
Note that orbitally stable limit cycles are also present in the unhatched blue area in Fig.~\ref{Fig2}a, where bistability between the petals and vortical states exists, and generally in the symmetry breaking state region of stability.
However, in this case the competition between the limit cycles and the symmetry breaking fixed points, mostly governed by the volumes of corresponding basins of attraction, complicates reliable realization of the space-time ordered state.

Linearization of Eq. \eqref{eq_AM} in the vicinity of the bifurcation $P_{AH}$ yields elliptic precession in the $yz$ plane with the frequency $\omega_0 = \delta P \sqrt{\varepsilon\left( 2 \xi - \varepsilon \right)}$.
For $P>P_{AH}$ it transforms into anharmonic periodic rotation of the pseudospin, characterized with frequency combs, shown in Fig.~\ref{Fig3}b.
Its inversed period $T^{-1}$, calculated as the main harmonic frequency of the Eq. \eqref{eq_AM} numerical solution, decreases from $\omega_0/2\pi$ to zero, as shown in Fig. \ref{Fig2}b.
\begin{figure}[t!]
\includegraphics[width=1.0\linewidth]{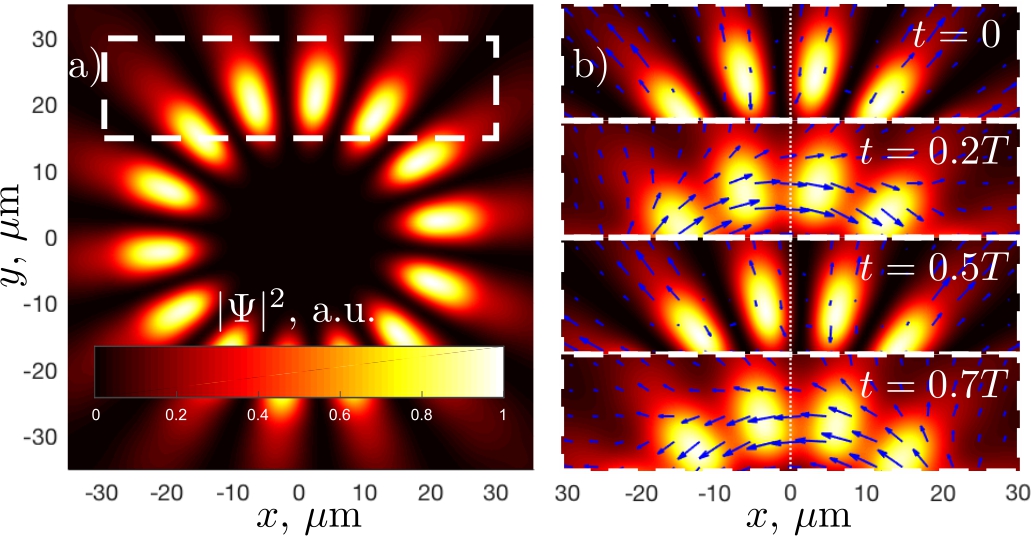}
\caption{(color online) Numerical solution of GPE in the time crystal regime. a) Polariton density time shot b) Evolution of instantaneous polariton density and current (shown with arrows) throughout the period $T$. \label{Fig4}}
\end{figure}

Fig.~\ref{Fig3}a shows the limit cycle trajectories in the pseudospin space, corresponding to pumping powers increasing within the instability range, and oscillations of the pseudospin projections, similar to the polarization pseudospin oscillations recently observed in the pulsed excitation regime \cite{Colas2015}.
The condensate density time shots, obtained from the full numerical solution of GPE \eqref{eq_GPE} and demonstrating its periodic evolution, are shown in Fig. \ref{Fig4}.
Numerics were performed using spectral methods in space and a variable-step, variable-order Adams-Bashforth-Moulton solver.
In order to avoid gain at the boundaries we use {\it ad hoc} finite pump shape introduced as $\mathrm{P}(\mathbf{r}) = P_0 \exp [ (r-r_0)/w ]^6$.
%\begin{equation}
%\mathrm{P}(\mathbf{r}) = P_0 \exp{\left[ \frac{r-r_0}{w}\right]^6}.
%\end{equation}
The parameters were set to: $\hbar m = 5 \times 10^{-5} m_0$ where $m_0$ is free electron mass, $\Gamma = 0.05 \ips$, $\gamma/\Gamma = 0.05$, $\beta = 5 \times 10^{-4} \ \ips \iium$, $\alpha/\beta = 1.6$, $\alpha_1/\beta = 6$, $P_0/\beta = 2000$, $r_0 = 25 \ \um$, and $w = 6 \ \um$. The wavefunction is initially seeded by stochastic white noise to mimic the incoherent uncondensed state.

The experimental conditions for the realization of the space-time crystal regime are twofold.
The spatial order is inherent for the linear limit of the GPE \eqref{eq_GPE} and requires $R\sqrt{2m\Gamma} > 1$ so that $\mathrm{m} > 1$, as follows from Fig.\ref{Fig1}b.
The temporal order in turn emerges in the nonlinear regime above the condensation threshold provided $\xi \approx \varepsilon$ (see Fig.\ref{Fig2}a).
Estimating $2 I_{cc} / (p_t^\mathrm{m} I _{cr}) \sim 10$ for $m \sim 10$, and $\gamma / \Gamma \sim 10$ \cite{Wouters2008b,Lagoudakis2008}, the condition on the interaction parameters transforms to the condition on the Hopfield coefficient $\vert X \vert^2 \sim 0.01$.

In contrast to existing realizations of space-time crystals, the periodicity of the condensate oscillations is governed by the optical trap parameters rather than the optical pumping frequency.
The inverse oscillation period scales from $\alpha \delta \mathrm{P} / \hbar $ to zero while the pumping power spans the limit cycle instability range (see Fig.\ref{Fig2}b), suggesting that the time crystal regime may be observed with time resolution of polariton emission by fine tuning of the optical trap parameters.

In conclusion, we predict a new space-time ordered phase of polariton condensates, created and trapped by excitonic reservoirs of annular shapes, which has the properties of a time crystal.
This phase arises from limit cycle instability in the vicinity of spontaneous parity breaking transition from petals to quantum vortices and can be described in terms of the condensate pseudospin rotation.
The physical origin of the emerging space-time order is in the interplay of strong interactions and driven-dissipative nature of exciton-polariton condensates.
%YR: In the conclusion paragraph, I replaced "condensate angular momentum rotation" by "condensate
%pseudospin rotation", and removed "nonlinear" from "strong nonlinear interactions". 

This work has been funded by Megagrant 14.Y26.31.0015 and Goszadanie no. 3.2614.2017/4.6 of the Ministry of Education and Science of Russian Federation and Icelandic Research Fund, Grant No. 163082-051.
A.N. acknowledges support from RFBR Grant No. 18-32-00434.
YGR acknowledges support from CONACYT (Mexico) under the Grant No.\ 251808.
We acknowledge A. Askitopoulos and P. Lagoudakis for stimulating discussions.

\bibliography{Mendeley}

\end{document}